\documentclass[twocolumn,preprintnumbers,amsmath,amssymb]{revtex4}

\usepackage{graphicx}
\usepackage{dcolumn}
\usepackage{bm,epsfig}
\usepackage{wasysym}
\usepackage{enumerate}
\usepackage{color}
\usepackage[breaklinks,colorlinks = true,linkcolor = red,urlcolor=cyan,citecolor=red]{hyperref}

\begin{document}

\title{Quantum oscillations in flux-grown SmB$_6$ with embedded aluminum}
\author{S. M. Thomas$^{1,2}$}
\author{Xiaxin Ding$^{3}$}
\author{F. Ronning$^{2}$}
\author{V. Zapf$^{3}$}
\author{J. D. Thompson$^{2}$}
\author{Z. Fisk$^{1}$}
\author{J. Xia$^{1}$}
\author{P. F. S. Rosa$^{2}$}
\affiliation{
$^{1}$ Department of Physics and Astronomy, University of California, Irvine, California 92967.\\
$^{2}$  Los Alamos National Laboratory, Los Alamos, New Mexico 87545, U.S.A.\\
$^{3}$ National High Magnetic Field Laboratory, Los Alamos, New Mexico 87545, U.S.A.}

\date{\today}

\begin{abstract}
SmB$_6$ is a candidate topological Kondo insulator that displays surface conduction at low temperatures.
Here, we perform torque magnetization measurements as a means to detect de Haas-van Alphen (dHvA) oscillations in SmB$_6$ crystals grown by aluminum flux.
We find that dHvA oscillations occur in single crystals containing embedded aluminum, originating from the flux used to synthesize SmB$_{6}$. 
Measurements on a sample with multiple, unconnected aluminum inclusions show that aluminum crystallizes in a preferred orientation within the SmB$_6$ cubic lattice.
The presence of aluminum is confirmed through bulk susceptibility measurements, but does not show a signature in transport measurements.
\end{abstract}

\maketitle

Single crystalline SmB$_6$ has been studied since the 1970s, but many mysteries still remain.
SmB$_6$ was initially viewed as a prototypical Kondo insulator,
in which incoherent scattering from $f$-electrons occurs at high temperatures whereas an insulating gap---driven by the hybridization between $f$ states and $d$ conduction bands---opens at low temperatures\cite{Fisk1996}.
A puzzling resistance saturation near 4~K was dismissed as arising from in-gap impurity states\cite{Menth1969},
but theoretical models recently suggested that SmB$_6$ is a topological Kondo insulator with conductive surface states and a robust bulk gap~\cite{Dzero2010,Takimoto2011}.
Thickness-dependent transport measurements in crystals grown via aluminum flux have shown that the resistance plateau is due to a metallic surface state surrounding the insulating 
bulk~\cite{Kim2013, Wolgast2013}.
Recent inverted Corbino measurements on the same crystals show that the bulk of SmB$_6$ displays a 10-order-of-magnitude increase in resistance with decreasing temperature, 
indicating that the bulk is truly insulating.\cite{Eo2018}
Nonetheless, SmB$_{6}$ grown by the floating zone method was claimed to host an exotic bulk Fermi surface (FS) in an insulating state.\cite{Tan2015}

Direct evidence of the expected topological helical structure of the surface states in SmB$_{6}$, however, remains elusive, and probes other than electrical
 transport are imperative.
Spin-dependent angle resolved photoemission spectroscopy (ARPES), which provides information on the band dispersion near the FS, was an obvious first choice.
ARPES experiments in SmB$_6$ have revealed in-gap states~\cite{Neupane2013,Xu2013,Jiang2013}, but issues with spin-resolved ARPES resolution compared to the small 
hybridization gap have made direct 
observation of spin-momentum locking in the surface states challenging~\cite{Neupane2013}.

Further information about the FS can be obtained through quantum oscillation measurements,
which provide information on the FS via angular dependent measurements of the extremal areas~\cite{Shoenberg1984}.
Although quantum oscillations have not been observed in the resistivity of SmB$_6$, two independent reports have been made on de Haas-van Alphen oscillations (dHvA, oscillations in the magnetization).
In the first report, dHvA oscillations in flux-grown crystals were attributed to a two-dimensional (2D) FS arising from the metallic surface state~\cite{Li2014}.
Contrary to claims of a heavy cyclotron mass observed in studies using thermopower and scanning tunneling spectroscopy\cite{Luo2015,Hamidian2018}, the cyclotron mass extracted from these dHvA measurements was found to be on the order of 0.1~$m_e$.
Considering the high mobility and light mass, it is remarkable that experimental evidence of oscillations has not been found in transport measurements.
Further, the origin of the surface state was thought to be the hybridization between the conduction band and the heavy Sm $f$-electrons, which also suggests a heavy surface 
state.
In the second report of quantum oscillations, the measured FS in floating-zone-grown crystals was claimed to have three-dimensional (3D) shape and to arise from the insulating 
bulk states~\cite{Tan2015}.
This result is also unexpected considering that quantum oscillations are traditionally observed in clean, metallic systems.
Nonetheless, five different theoretical explanations have been reported for both the light electrons observed in the 2D FS~\cite{Alexandrov2015}, and for the presence of 
oscillations arising from an insulating state~\cite{Knolle2015, Erten2016, Pal2018, Shen2018}.

To shed light on this controversy, here we use torque magnetometry to measure quantum oscillations in the magnetization of flux-grown SmB$_{6}$ as a function of its thickness.
We find that flux-grown crystals only exhibit dHvA oscillations when embedded aluminum is present.
The Al inclusions co-crystallize with the SmB$_6$ host crystal, with the [100] Al axis nearly aligned with the [100] SmB$_6$ axis.
Angular dependence of our dHvA oscillations is in good agreement with those reported previously for single crystalline Al~\cite{Larson1967}.
Interestingly, Al inclusions in the crystal bulk show no evidence for a superconducting transition in transport measurements.
We note that the results reported here have no correlation with the observations of the 3D FS reported by Tan \textit{et al.}, and should not be taken 
as a proof (or disproof) of those observations~\cite{Tan2015}.

For our investigation, we choose single crystals of SmB$_{6}$ grown using the aluminum flux technique.
More information on this technique can be found in the supplemental information.
The inset of Fig.~1a shows a typical flux-grown crystal of SmB$_6$ with dimensions $3\times 2\times 1$~mm$^{3}$.
Aluminum does not substitute into the hexaboride lattice, but larger crystals often enclose Al pockets which can be mechanically removed by polishing or chemically etched with hydrochloric acid.
We note that Al also crystallizes in a cubic space group, \textit{Fm-3m} (225), with a lattice parameter $a=4.05$~\AA$\,$ that is only 2\% smaller than that of SmB$_6$.

\begin{figure}[!h]
\begin{center}
\includegraphics[width=1.0\columnwidth]{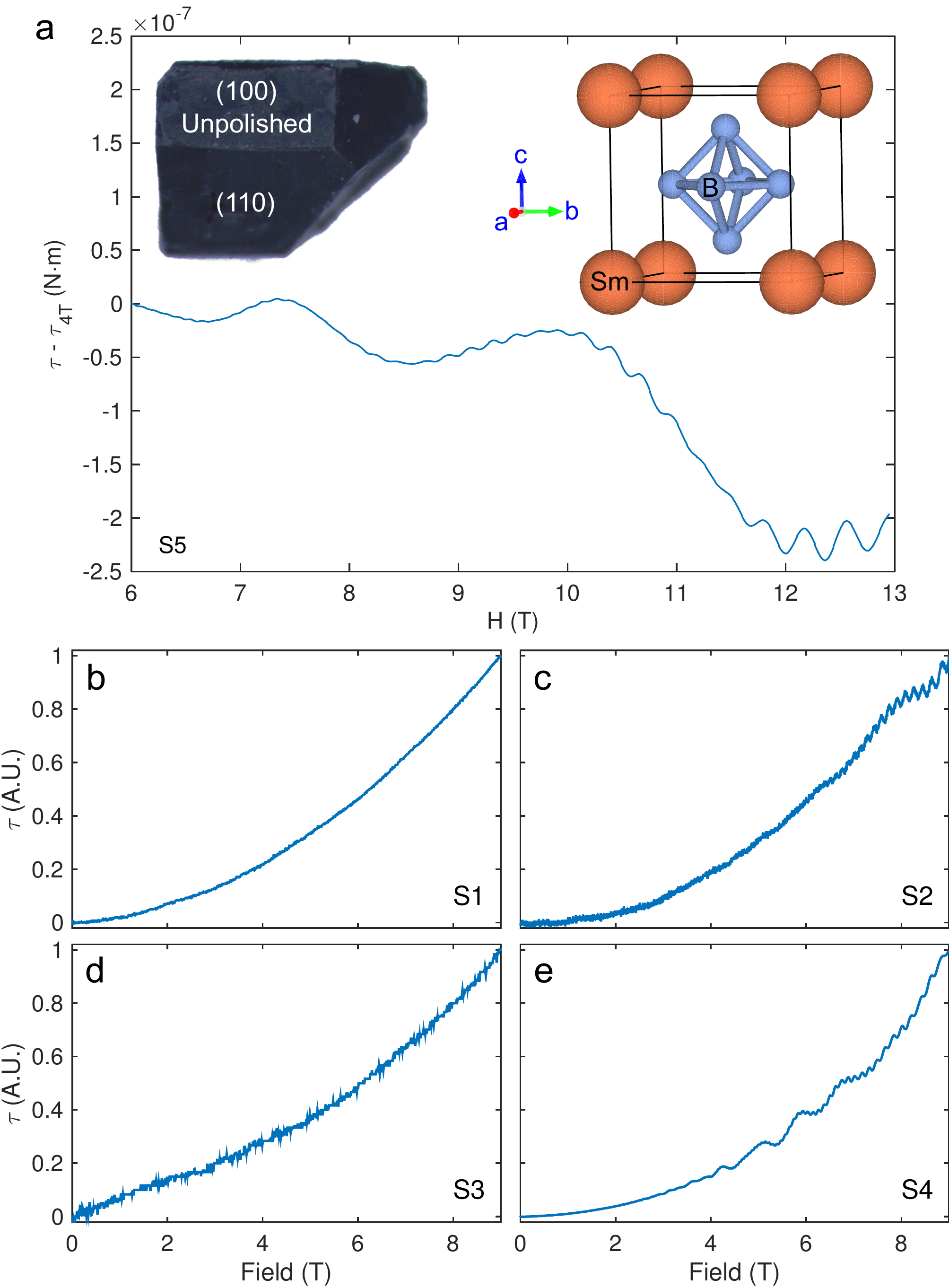}
\vspace{-0.7cm}
\end{center}
\caption{
(a)~Torque magnetization as a function of magnetic field of a representative SmB$_{6}$ single crystal (s5).
To focus on the oscillations, the value of the torque at 4~T was subtracted from the data.
The inset shows a picture of the as-grown sample along with its cubic crystal structure.
(b-e)~A survey of several SmB$_{6}$ crystals that were checked for quantum oscillations using torque magnetometry. Only samples s2 and s4 showed oscillations.
}\label{fig:oscillations}
\end{figure}

Quantum oscillations arise in many physical properties of metallic materials under the condition that $\omega_c\tau > 1$,
where $\omega_c$ is the cyclotron frequency and $\tau$ is the electron scattering time.
Onsager showed that the oscillation period in inverse field is proportional to the cross sectional area of the FS~\cite{Onsager1952}:
\begin{equation}
\Delta{}\bigg(\frac{1}{B}\bigg)=\frac{2\pi{}e}{\hbar{}}\frac{1}{A_{e}}
\end{equation}
For a 2D material, the FS is expected to have cylindrical character.
The oscillation frequency for a [001] rotation axis should vary as $1/\cos(\theta-\phi_S)$,
where $\theta$ is angle between [100] and the field and $\phi_S$ is the angle between a surface normal and [100].
Because SmB$_6$ crystals grow with (100) and (110) facets, surface states on these facets should have $\phi_S=90 n$ and $\phi_S=45+90n$ degrees, respectively, where $n$ is an integer.
In a 3D material, the frequency will also diverge along any open orbits on the FS.

In an attempt to determine the nature of the quantum oscillations in SmB$_6$, torque magnetometry measurements were performed on many single crystals.
As shown in Fig.~1b--e, however, only a subset of SmB$_6$ samples showed oscillatory behavior in magnetization, and these samples tended to have larger thickness.
The lack of oscillations in some of the samples, despite having similar surface facets and surface area, was the first indication that the presence of oscillations may not be 
intrinsic to the surface state of SmB$_6$ crystals.
As shown in the Supplemental Information, magnetoresistance at 50~mK was also measured in one of the samples that showed dHvA oscillations.
After subtracting a polynomial background, the frequency content of the magnetoresistance was calculated.
The lack of any clear peak in the frequency spectrum shows that Shubnikov-de Haas (SdH) oscillations are not detectable in fields up to 12~T, even at 50~mK.
This result is consistent with a recent magnetoresistance study on SmB$_6$ at temperatures as low as 300~mK using special contact structures to only measure the contribution for individual crystal surfaces~\cite{Wolgast2015}.

\begin{figure}[!ht]
\begin{center}
\includegraphics[width=.9\columnwidth]{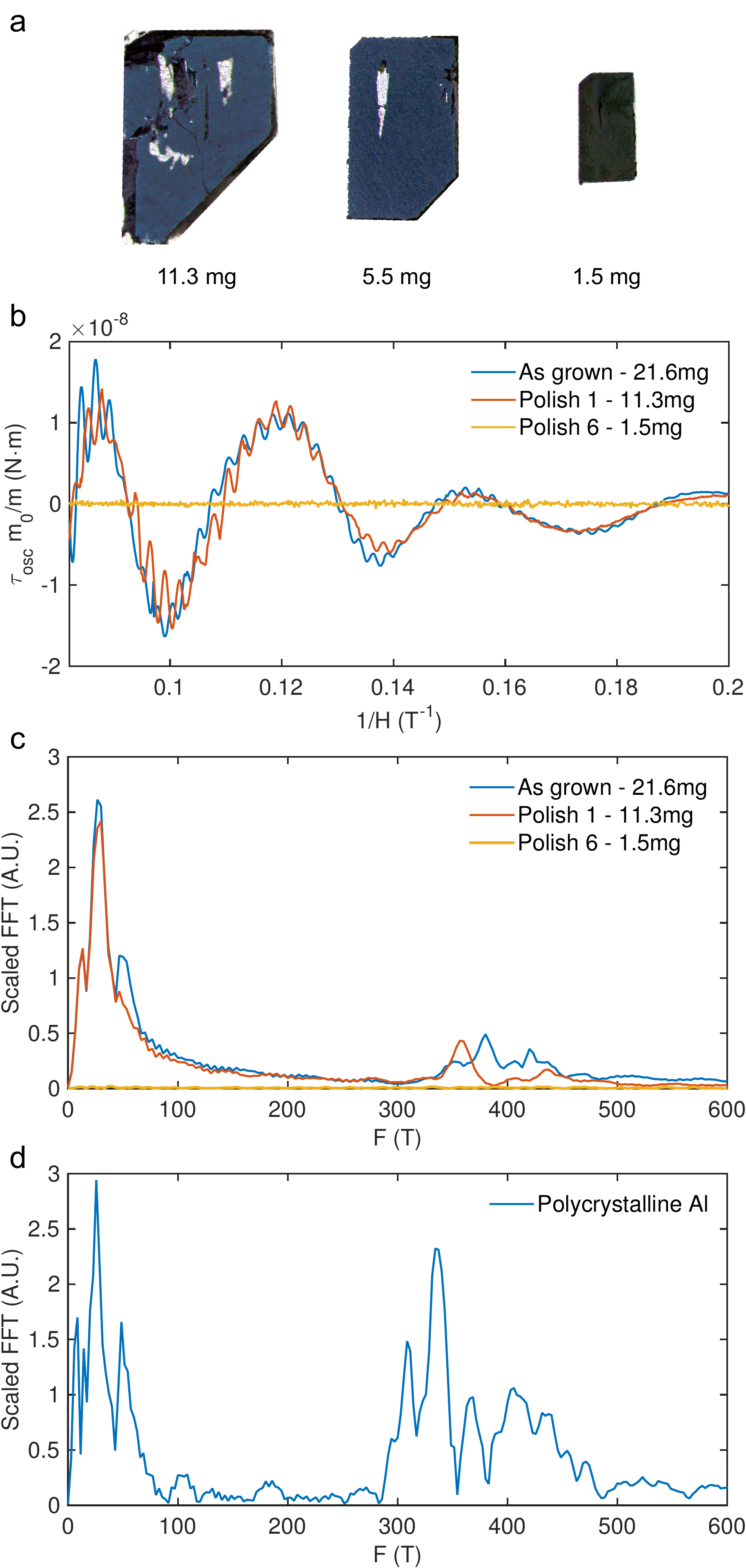}
\vspace{-0.7cm}
\end{center}
\caption{
(a)~Progressive polishing of sample s5, showing a series of aluminum deposits distributed throughout the sample. The initial (100) surface was left undisturbed during the polishing process.
(b)~Oscillatory torque versus inverse field for s5 at several different polishing steps.
The torque is scaled with the mass of the sample, and field is applied a few degrees from [010].
After the last Aluminum deposit is removed, the oscillations vanish.
(c)~Scaled frequency spectra of the oscillatory torque shown above.
(d)~Frequency spectrum of polycrystalline Al flux. There is broad spectral weight between 300 and 500~T, in contrast to the oriented single-crystal Aluminum in SmB$_6$.
}\label{fig:al_inclusion}
\end{figure}

One of the crystals exhibiting dHvA oscillations (s5, Fig.~1a) was polished to determine whether the oscillations are thickness dependent.
Only the bottom surface was polished, and care was taken to keep the top surface shown in Fig.~1a intact.
After each polishing step, any exposed aluminum was etched away using hydrochloric acid.
As shown in Fig.~2a (left panel), three disconnected aluminum inclusions appeared after polishing away the bottom portion of the crystal.
After further polishing, several more disconnected inclusions were discovered, one of which is shown in the middle panel of Fig.~2a.
At the end, the sample was polished to $230$ microns and no Al inclusions were apparent (Fig.~2a, right panel).

Fig.~2b shows the torque magnetization obtained in the as grown sample ($m = 21.6$~mg) compared to the signal obtained after polishing away roughly half of the sample  ($m = 11.3$~mg).
Remarkably, frequency analysis shown in Fig.~2c revealed that the FFT amplitude roughly scales with the mass of the sample and not the sample area---
consistent with oscillations arising from Al inclusions that are homogeneously distributed in the bulk of the crystal.
Moreover, well-defined peaks exhibiting clear angular dependence are observed in the frequency spectrum, despite the presence of multiple aluminum inclusions.

In contrast, Fig.~2d shows dHvA oscillations from a small piece of 5N aluminum used as flux during the growth process.
There are more than five peaks in the 300--500~T range due to the fact that the pellet is composed of many randomly oriented microcrystals.
Further, polycrystalline Al does not exhibit a clear pattern in angle-dependent measurements.\cite{Li2014}
Considering the multitude of distinct aluminum inclusions in this particular SmB$_6$ crystal,
the relatively sparse spectrum with well-defined angular dependence shows that the inclusions are preferentially aligned along the same crystallographic axis.
The fact that embedded aluminum inclusions co-crystallize with the SmB$_6$ was also reported in a study combining neutron diffraction, powder diffraction, and x-ray computed tomography.\cite{Phelan2016a}

To check that the sole source of the oscillations was embedded aluminum deposits, the sample was polished to a thin plate as shown in the right panel of Fig.~2a.
As shown in Figs.~2b--c, there are no oscillations observed after the final polish and etching step, even though the (100) surface on the top of the sample has been left undisturbed.
This confirms that the source of the observed oscillations is the embedded Al deposits in SmB$_6$.

\begin{figure}[b]
\begin{center}
\includegraphics[width=.9\columnwidth]{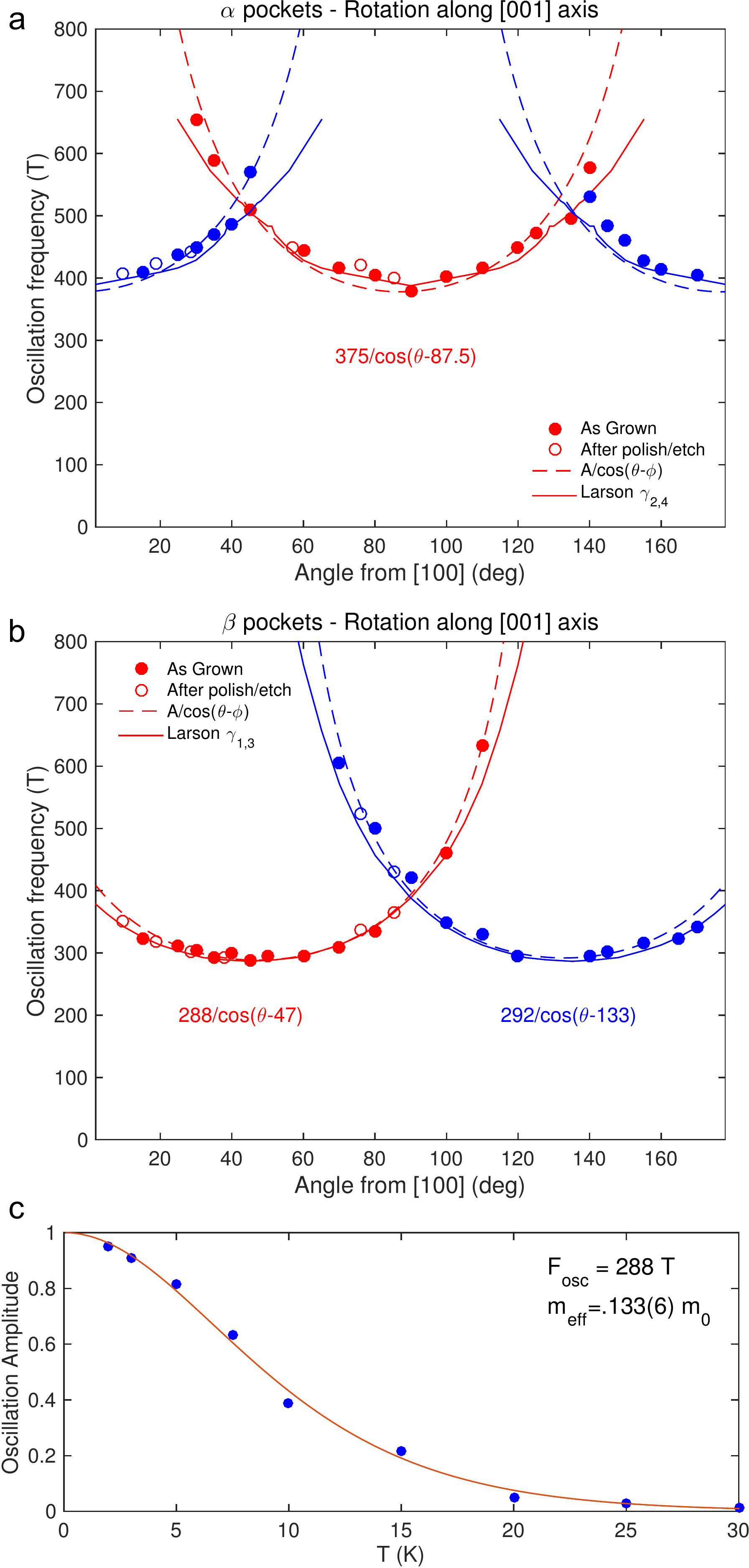}
\vspace{-0.7cm}
\end{center}
\caption{
(a) Angular dependence of the dHvA oscillation frequency for the $\alpha$ pocket.
(b) Angular dependence of the dHvA oscillation frequency for the $\beta$ pocket.
(c) Thermal damping of the $\beta$ oscillation at 45$^\circ$ from [100].
}\label{fig:angular_dependence}
\end{figure}

Having established the origin of the dHvA signal in SmB$_6$, we now briefly turn to the temperature- and angle-dependence of the oscillation frequencies.
The magnitude of the dHvA oscillations follows the temperature dependence given by the Lifshitz-Kosevich formula:\cite{Shoenberg1984}
\begin{equation}
R_{T}=\alpha{}Tm^*/B\sinh{}(\alpha{}Tm^*/B),
\end{equation}
\noindent and Fig.~3c shows a fit of the 288~T oscillation amplitude at 8.05~T applied 45$^\circ$ from [100] to the thermal damping equation, which gives an effective mass of 0.133~m$_e$.
This angle was chosen because it provides the largest separation of the oscillations in frequency.
The value for the effective mass agrees with a previous report on single-crystal aluminum that found an effective mass of 0.130(4)~m$_e$ when field was applied along the same direction.\cite{Larson1967}

Fig.~3a--b shows the angular dependence of the dHvA oscillations in SmB$_6$ as the crystal is rotated about the [001] axis.
The pockets with minimum frequency near 375~T were assigned the $\alpha$-pocket designation, whereas those with minimum frequency near 290~T were designated the $\beta$ pockets.
The observed oscillation frequencies compare well with those for single-crystal aluminum as reported by Larson \textit{et al.}\cite{Gunnersen1957,Larson1967}
Larson measured single-crystal Al from [010] to [110] rotating in the (100) plane and assigned four pocket designations ($\gamma_{1-4}$) for frequencies in the 200--1000~T range.
Because Al is four-fold symmetric in the (100) plane, these designations are actually two pockets that repeat every 90 degrees.  
$\gamma_1$ and $\gamma_3$ correspond to a pocket with minimum oscillation near 285~T, and $\gamma_2$ and $\gamma_4$ correspond to a pocket with minimum oscillation near 390~T.
This remarkable similarity further confirms our scenario that the [001] axis of the Aluminum inclusions in SmB$_6$ is very nearly aligned with the SmB$_6$ [001] axis.

The small difference at larger angles may be attributed to the presence of small amounts of strain due to the 0.08 \AA{} mismatch in lattice parameters between Al and SmB$_6$.
Fits to the expected angular dependence of a 2D FS, $F_0/\cos(\theta-\phi_S)$, are also shown for comparison.
Lastly, the angular dependence was also measured after etching the three Aluminum deposits depicted in Fig.~2a (left panel).
Removing nearly half of the embedded aluminum had little effect on the angular dependence of the observed oscillations as shown by the open symbols in Fig.~3a--b.
Again, this demonstrates that the Al inclusions are co-aligned.

Aluminum has a superconducting critical temperature of 1.17 K and a critical field of 105~Oersted.\cite{Caplan1965}
Fig.~4a shows transport measurements performed on an SmB$_6$ crystal near the superconducting transition of Al.
Remarkably, no feature is visible in resistivity.
Down to 2~K, the bulk of SmB$_6$ is truly insulating as shown recently by Y. S. Eo \textit{et al.},
which explains the lack of SdH oscillations in SmB$_6$ or transport evidence of the superconducting transition from subsurface Al inclusions.\cite{Eo2018}
In contrast, the presence of aluminum can be detected through bulk magnetization measurements.
As shown in Fig.~4b, evidence of the aluminum superconducting transition is visible in AC susceptibility measurements.
Above 1.2~K, no feature is observed in the susceptibility data,
but as the temperature is lowered, a feature emerges and tracks the critical field expected for the Al superconducting transition.

\begin{figure}[]
    \begin{center}
    \includegraphics[width=.9\columnwidth]{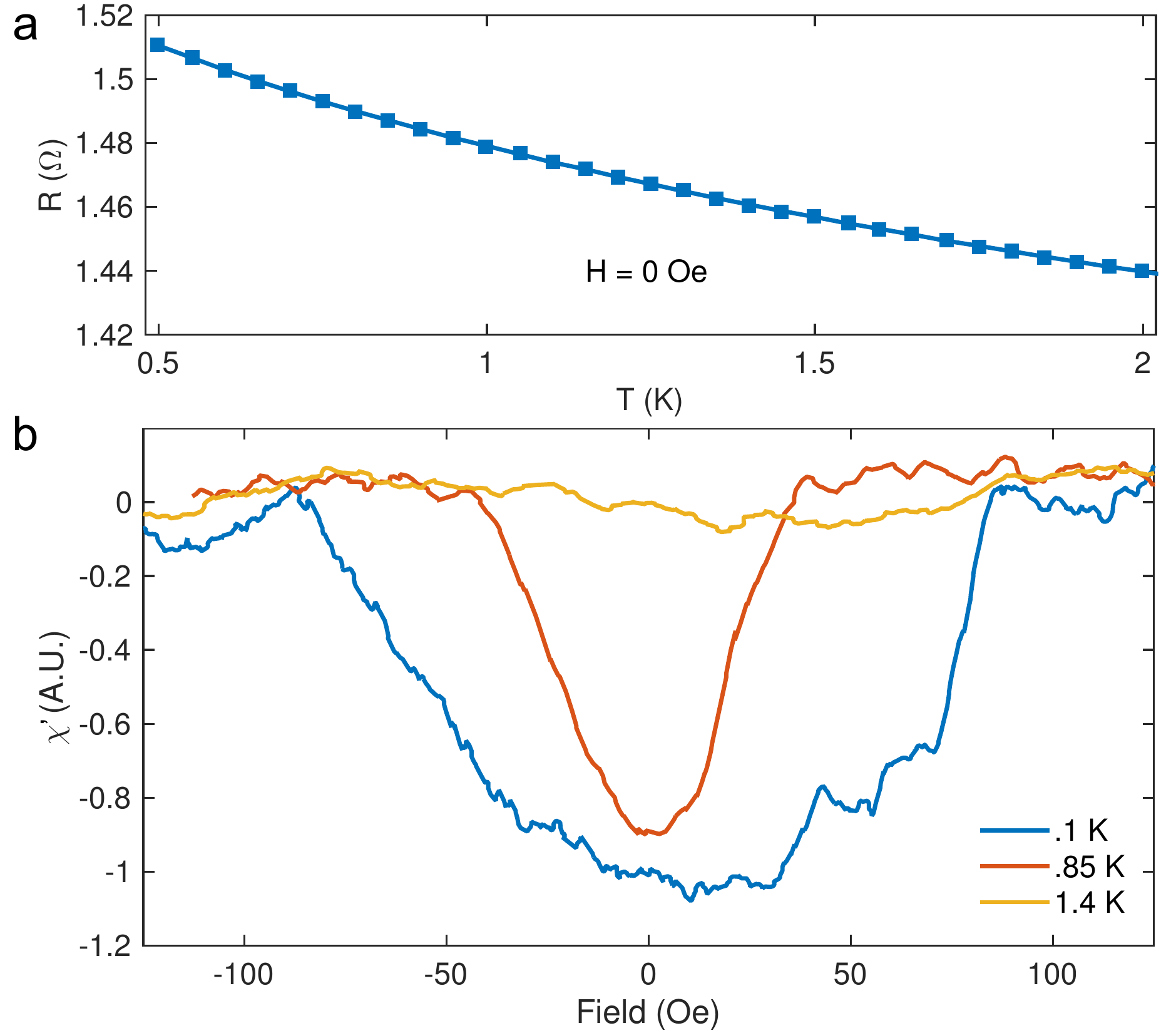}
    \vspace{-0.7cm}
    \end{center}
    \caption{
    (a) Resistance versus temperature in zero applied field for a SmB$_6$ crystal with a subsurface Aluminum inclusion.
    (b) AC Susceptibility measurement of SmB$_6$ crystal near zero field.
    }\label{fig:al_detection}
\end{figure}

After these transport and susceptibility measurements, the sample was polished to determine the proximity of the Al inclusions to the surface.
Subsurface Al deposits became visible after only a few polishing laps, showing that the inclusion was separated from the surface by less than 100~$\mu$m.
Due to the highly insulating bulk in SmB$_{6}$ at low temperatures, an aluminum inclusion that is shallowly embedded within the bulk is completely isolated from the metallic surface state.
Thus, when screening SmB$_6$ samples for aluminum inclusions, resistance measurements are insufficient.

It should be noted that the study on the 3D FS reported by Tan \textit{et al.} was performed on SmB$_6$ crystals grown by the floating-zone method.\cite{Tan2015}
There is growing evidence that different growth methods yield SmB$_{6}$ crystals with non-negligible differences.\cite{Phelan2016a,Valentine2016,YSEoMM2018}
Further, Tan \textit{et al.} observed a large departure from the Lifshitz-Kosevich temperature dependence for oscillation amplitudes below 500~mK.
We have measured the oscillation amplitude of flux-grown SmB$_{6}$ down to 50~mK (see Supplemental Information), and have observed no departure from the Lifshitz-Kosevich behavior.
We do not observe the same angular dependence or frequency spectrum as Tan \textit{et al.}, and it remains to be verified how the synthesis technique influences these results.
This shows that the results reported here have no correlation with the observations of the 3D FS reported by Tan \textit{et al.}, and should not be taken as a proof (or disproof) of those observations.\cite{Tan2015}

In conclusion, we have shown that dHvA oscillations in flux-grown SmB$_6$ can arise from subsurface Aluminum inclusions.
The inclusions are nearly aligned with the SmB$_6$ [001] crystal axis and provide quantum oscillations with an effective mass of 0.1~m$_e$.
After completely removing all aluminum inclusions, the dHvA oscillation signal vanishes.
Angular dependence shows that the orbits are in good agreement with those of single crystalline Al reported by Larson \textit{et al}.\cite{Larson1967}
Our results demonstrate that, when performing measurements on SmB$_6$ crystals, it is necessary to screen the samples for aluminum by using a bulk technique capable of probing beyond the metallic surface state.

\begin{acknowledgments}
We would like to acknowledge constructive discussions with C. Kurdak and Y. S. Eo.  
Work at Los Alamos was supported by the Laboratory Directed Research and Development program under project number 20160085DR.
Work at UCI was supported by NSF grant No.~DMR-1708199.  
A portion of this work was performed at the National High Magnetic Field Laboratory,
which is supported by National Science Foundation Cooperative Agreement No.~DMR-1157490.
\end{acknowledgments}


\begin{thebibliography}{28}
    \expandafter\ifx\csname natexlab\endcsname\relax\def\natexlab#1{#1}\fi
    \expandafter\ifx\csname bibnamefont\endcsname\relax
      \def\bibnamefont#1{#1}\fi
    \expandafter\ifx\csname bibfnamefont\endcsname\relax
      \def\bibfnamefont#1{#1}\fi
    \expandafter\ifx\csname citenamefont\endcsname\relax
      \def\citenamefont#1{#1}\fi
    \expandafter\ifx\csname url\endcsname\relax
      \def\url#1{\texttt{#1}}\fi
    \expandafter\ifx\csname urlprefix\endcsname\relax\def\urlprefix{URL }\fi
    \providecommand{\bibinfo}[2]{#2}
    \providecommand{\eprint}[2][]{\url{#2}}
    
    \bibitem[{\citenamefont{Fisk et~al.}(1996)\citenamefont{Fisk, Sarrao, Cooper,
      Nyhus, Boebinger, Passner, and Canfield}}]{Fisk1996}
    \bibinfo{author}{\bibfnamefont{Z.}~\bibnamefont{Fisk}},
      \bibinfo{author}{\bibfnamefont{J.}~\bibnamefont{Sarrao}},
      \bibinfo{author}{\bibfnamefont{S.}~\bibnamefont{Cooper}},
      \bibinfo{author}{\bibfnamefont{P.}~\bibnamefont{Nyhus}},
      \bibinfo{author}{\bibfnamefont{G.}~\bibnamefont{Boebinger}},
      \bibinfo{author}{\bibfnamefont{A.}~\bibnamefont{Passner}}, \bibnamefont{and}
      \bibinfo{author}{\bibfnamefont{P.}~\bibnamefont{Canfield}},
      \bibinfo{journal}{Physica B: Condensed Matter}
      \textbf{\bibinfo{volume}{223-224}}, \bibinfo{pages}{409}
      (\bibinfo{year}{1996}), ISSN \bibinfo{issn}{09214526},
      \urlprefix\url{http://linkinghub.elsevier.com/retrieve/pii/0921452696001366}.
    
    \bibitem[{\citenamefont{Menth et~al.}(1969)\citenamefont{Menth, Buehler, and
      Geballe}}]{Menth1969}
    \bibinfo{author}{\bibfnamefont{A.}~\bibnamefont{Menth}},
      \bibinfo{author}{\bibfnamefont{E.}~\bibnamefont{Buehler}}, \bibnamefont{and}
      \bibinfo{author}{\bibfnamefont{T.~H.} \bibnamefont{Geballe}},
      \bibinfo{journal}{Physical Review Letters} \textbf{\bibinfo{volume}{22}},
      \bibinfo{pages}{295} (\bibinfo{year}{1969}), ISSN \bibinfo{issn}{0031-9007},
      \urlprefix\url{https://link.aps.org/doi/10.1103/PhysRevLett.22.295}.
    
    \bibitem[{\citenamefont{Dzero et~al.}(2010)\citenamefont{Dzero, Sun, Galitski,
      and Coleman}}]{Dzero2010}
    \bibinfo{author}{\bibfnamefont{M.}~\bibnamefont{Dzero}},
      \bibinfo{author}{\bibfnamefont{K.}~\bibnamefont{Sun}},
      \bibinfo{author}{\bibfnamefont{V.}~\bibnamefont{Galitski}}, \bibnamefont{and}
      \bibinfo{author}{\bibfnamefont{P.}~\bibnamefont{Coleman}},
      \bibinfo{journal}{Physical Review Letters} \textbf{\bibinfo{volume}{104}},
      \bibinfo{pages}{106408} (\bibinfo{year}{2010}), ISSN
      \bibinfo{issn}{0031-9007}, \eprint{0912.3750},
      \urlprefix\url{http://arxiv.org/abs/0912.3750{\%}0Ahttp://dx.doi.org/10.1103/PhysRevLett.104.106408
      https://link.aps.org/doi/10.1103/PhysRevLett.104.106408}.
    
    \bibitem[{\citenamefont{Takimoto}(2011)}]{Takimoto2011}
    \bibinfo{author}{\bibfnamefont{T.}~\bibnamefont{Takimoto}},
      \bibinfo{journal}{Journal of the Physical Society of Japan}
      \textbf{\bibinfo{volume}{80}}, \bibinfo{pages}{123710}
      (\bibinfo{year}{2011}), ISSN \bibinfo{issn}{0031-9015},
      \urlprefix\url{http://journals.jps.jp/doi/10.1143/JPSJ.80.123710}.
    
    \bibitem[{\citenamefont{Kim et~al.}(2013)\citenamefont{Kim, Thomas, Grant,
      Botimer, Fisk, and Xia}}]{Kim2013}
    \bibinfo{author}{\bibfnamefont{D.~J.} \bibnamefont{Kim}},
      \bibinfo{author}{\bibfnamefont{S.}~\bibnamefont{Thomas}},
      \bibinfo{author}{\bibfnamefont{T.}~\bibnamefont{Grant}},
      \bibinfo{author}{\bibfnamefont{J.}~\bibnamefont{Botimer}},
      \bibinfo{author}{\bibfnamefont{Z.}~\bibnamefont{Fisk}}, \bibnamefont{and}
      \bibinfo{author}{\bibfnamefont{J.}~\bibnamefont{Xia}},
      \bibinfo{journal}{Scientific Reports} \textbf{\bibinfo{volume}{3}},
      \bibinfo{pages}{3150} (\bibinfo{year}{2013}), ISSN \bibinfo{issn}{2045-2322},
      \eprint{1211.6769}, \urlprefix\url{http://www.nature.com/articles/srep03150}.
    
    \bibitem[{\citenamefont{Wolgast et~al.}(2013)\citenamefont{Wolgast, Kurdak,
      Sun, Allen, Kim, and Fisk}}]{Wolgast2013}
    \bibinfo{author}{\bibfnamefont{S.}~\bibnamefont{Wolgast}},
      \bibinfo{author}{\bibfnamefont{{\c{C}}.}~\bibnamefont{Kurdak}},
      \bibinfo{author}{\bibfnamefont{K.}~\bibnamefont{Sun}},
      \bibinfo{author}{\bibfnamefont{J.~W.} \bibnamefont{Allen}},
      \bibinfo{author}{\bibfnamefont{D.-J.} \bibnamefont{Kim}}, \bibnamefont{and}
      \bibinfo{author}{\bibfnamefont{Z.}~\bibnamefont{Fisk}},
      \bibinfo{journal}{Physical Review B} \textbf{\bibinfo{volume}{88}},
      \bibinfo{pages}{180405} (\bibinfo{year}{2013}), ISSN
      \bibinfo{issn}{1098-0121}, \eprint{1211.5104},
      \urlprefix\url{https://link.aps.org/doi/10.1103/PhysRevB.88.180405}.
    
    \bibitem[{\citenamefont{Eo et~al.}(2018{\natexlab{a}})\citenamefont{Eo,
      Rakoski, Lucien, Mihaliov, Kurdak, Rosa, Kim, and Fisk}}]{Eo2018}
    \bibinfo{author}{\bibfnamefont{Y.~S.} \bibnamefont{Eo}},
      \bibinfo{author}{\bibfnamefont{A.}~\bibnamefont{Rakoski}},
      \bibinfo{author}{\bibfnamefont{J.}~\bibnamefont{Lucien}},
      \bibinfo{author}{\bibfnamefont{D.}~\bibnamefont{Mihaliov}},
      \bibinfo{author}{\bibfnamefont{C.}~\bibnamefont{Kurdak}},
      \bibinfo{author}{\bibfnamefont{P.~F.~S.} \bibnamefont{Rosa}},
      \bibinfo{author}{\bibfnamefont{D.-J.} \bibnamefont{Kim}}, \bibnamefont{and}
      \bibinfo{author}{\bibfnamefont{Z.}~\bibnamefont{Fisk}}
      (\bibinfo{year}{2018}{\natexlab{a}}), \eprint{1803.00959},
      \urlprefix\url{http://arxiv.org/abs/1803.00959}.
    
    \bibitem[{\citenamefont{Tan et~al.}(2015)\citenamefont{Tan, Hsu, Zeng, Hatnean,
      Harrison, Zhu, Hartstein, Kiourlappou, Srivastava, Johannes
      et~al.}}]{Tan2015}
    \bibinfo{author}{\bibfnamefont{B.~S.} \bibnamefont{Tan}},
      \bibinfo{author}{\bibfnamefont{Y.-T.} \bibnamefont{Hsu}},
      \bibinfo{author}{\bibfnamefont{B.}~\bibnamefont{Zeng}},
      \bibinfo{author}{\bibfnamefont{M.~C.} \bibnamefont{Hatnean}},
      \bibinfo{author}{\bibfnamefont{N.}~\bibnamefont{Harrison}},
      \bibinfo{author}{\bibfnamefont{Z.}~\bibnamefont{Zhu}},
      \bibinfo{author}{\bibfnamefont{M.}~\bibnamefont{Hartstein}},
      \bibinfo{author}{\bibfnamefont{M.}~\bibnamefont{Kiourlappou}},
      \bibinfo{author}{\bibfnamefont{A.}~\bibnamefont{Srivastava}},
      \bibinfo{author}{\bibfnamefont{M.~D.} \bibnamefont{Johannes}},
      \bibnamefont{et~al.}, \bibinfo{journal}{Science}
      \textbf{\bibinfo{volume}{349}}, \bibinfo{pages}{287} (\bibinfo{year}{2015}),
      ISSN \bibinfo{issn}{0036-8075},
      \urlprefix\url{http://science.sciencemag.org/content/349/6245/287.abstract
      http://www.sciencemag.org/cgi/doi/10.1126/science.aaa7974}.
    
    \bibitem[{\citenamefont{Neupane et~al.}(2013)\citenamefont{Neupane, Alidoust,
      Xu, Kondo, Ishida, Kim, Liu, Belopolski, Jo, Chang et~al.}}]{Neupane2013}
    \bibinfo{author}{\bibfnamefont{M.}~\bibnamefont{Neupane}},
      \bibinfo{author}{\bibfnamefont{N.}~\bibnamefont{Alidoust}},
      \bibinfo{author}{\bibfnamefont{S.-Y.} \bibnamefont{Xu}},
      \bibinfo{author}{\bibfnamefont{T.}~\bibnamefont{Kondo}},
      \bibinfo{author}{\bibfnamefont{Y.}~\bibnamefont{Ishida}},
      \bibinfo{author}{\bibfnamefont{D.~J.} \bibnamefont{Kim}},
      \bibinfo{author}{\bibfnamefont{C.}~\bibnamefont{Liu}},
      \bibinfo{author}{\bibfnamefont{I.}~\bibnamefont{Belopolski}},
      \bibinfo{author}{\bibfnamefont{Y.~J.} \bibnamefont{Jo}},
      \bibinfo{author}{\bibfnamefont{T.-R.} \bibnamefont{Chang}},
      \bibnamefont{et~al.}, \bibinfo{journal}{Nature Communications}
      \textbf{\bibinfo{volume}{4}}, \bibinfo{pages}{2991} (\bibinfo{year}{2013}),
      ISSN \bibinfo{issn}{2041-1723}, \eprint{1312.1979},
      \urlprefix\url{http://www.nature.com/articles/ncomms3991}.
    
    \bibitem[{\citenamefont{Xu et~al.}(2013)\citenamefont{Xu, Shi, Biswas, Matt,
      Dhaka, Huang, Plumb, Radovi{\'{c}}, Dil, Pomjakushina et~al.}}]{Xu2013}
    \bibinfo{author}{\bibfnamefont{N.}~\bibnamefont{Xu}},
      \bibinfo{author}{\bibfnamefont{X.}~\bibnamefont{Shi}},
      \bibinfo{author}{\bibfnamefont{P.~K.} \bibnamefont{Biswas}},
      \bibinfo{author}{\bibfnamefont{C.~E.} \bibnamefont{Matt}},
      \bibinfo{author}{\bibfnamefont{R.~S.} \bibnamefont{Dhaka}},
      \bibinfo{author}{\bibfnamefont{Y.}~\bibnamefont{Huang}},
      \bibinfo{author}{\bibfnamefont{N.~C.} \bibnamefont{Plumb}},
      \bibinfo{author}{\bibfnamefont{M.}~\bibnamefont{Radovi{\'{c}}}},
      \bibinfo{author}{\bibfnamefont{J.~H.} \bibnamefont{Dil}},
      \bibinfo{author}{\bibfnamefont{E.}~\bibnamefont{Pomjakushina}},
      \bibnamefont{et~al.}, \bibinfo{journal}{Physical Review B}
      \textbf{\bibinfo{volume}{88}}, \bibinfo{pages}{121102}
      (\bibinfo{year}{2013}), ISSN \bibinfo{issn}{1098-0121}, \eprint{1306.3678},
      \urlprefix\url{https://link.aps.org/doi/10.1103/PhysRevB.88.121102}.
    
    \bibitem[{\citenamefont{Jiang et~al.}(2013)\citenamefont{Jiang, Li, Zhang, Sun,
      Chen, Ye, Xu, Ge, Tan, Niu et~al.}}]{Jiang2013}
    \bibinfo{author}{\bibfnamefont{J.}~\bibnamefont{Jiang}},
      \bibinfo{author}{\bibfnamefont{S.}~\bibnamefont{Li}},
      \bibinfo{author}{\bibfnamefont{T.}~\bibnamefont{Zhang}},
      \bibinfo{author}{\bibfnamefont{Z.}~\bibnamefont{Sun}},
      \bibinfo{author}{\bibfnamefont{F.}~\bibnamefont{Chen}},
      \bibinfo{author}{\bibfnamefont{Z.}~\bibnamefont{Ye}},
      \bibinfo{author}{\bibfnamefont{M.}~\bibnamefont{Xu}},
      \bibinfo{author}{\bibfnamefont{Q.}~\bibnamefont{Ge}},
      \bibinfo{author}{\bibfnamefont{S.}~\bibnamefont{Tan}},
      \bibinfo{author}{\bibfnamefont{X.}~\bibnamefont{Niu}}, \bibnamefont{et~al.},
      \bibinfo{journal}{Nature Communications} \textbf{\bibinfo{volume}{4}},
      \bibinfo{pages}{3010} (\bibinfo{year}{2013}), ISSN \bibinfo{issn}{2041-1723},
      \eprint{1306.5664},
      \urlprefix\url{http://www.nature.com/articles/ncomms4010}.
    
    \bibitem[{\citenamefont{Shoenberg}(1984)}]{Shoenberg1984}
    \bibinfo{author}{\bibfnamefont{D.}~\bibnamefont{Shoenberg}},
      \emph{\bibinfo{title}{{Magnetic oscillations in metals}}}
      (\bibinfo{publisher}{Cambridge University Press},
      \bibinfo{address}{Cambridge}, \bibinfo{year}{1984}), ISBN
      \bibinfo{isbn}{9780511897870},
      \urlprefix\url{http://ebooks.cambridge.org/ref/id/CBO9780511897870}.
    
    \bibitem[{\citenamefont{Li et~al.}(2014)\citenamefont{Li, Xiang, Yu, Asaba,
      Lawson, Cai, Tinsman, Berkley, Wolgast, Eo et~al.}}]{Li2014}
    \bibinfo{author}{\bibfnamefont{G.}~\bibnamefont{Li}},
      \bibinfo{author}{\bibfnamefont{Z.}~\bibnamefont{Xiang}},
      \bibinfo{author}{\bibfnamefont{F.}~\bibnamefont{Yu}},
      \bibinfo{author}{\bibfnamefont{T.}~\bibnamefont{Asaba}},
      \bibinfo{author}{\bibfnamefont{B.}~\bibnamefont{Lawson}},
      \bibinfo{author}{\bibfnamefont{P.}~\bibnamefont{Cai}},
      \bibinfo{author}{\bibfnamefont{C.}~\bibnamefont{Tinsman}},
      \bibinfo{author}{\bibfnamefont{A.}~\bibnamefont{Berkley}},
      \bibinfo{author}{\bibfnamefont{S.}~\bibnamefont{Wolgast}},
      \bibinfo{author}{\bibfnamefont{Y.~S.} \bibnamefont{Eo}},
      \bibnamefont{et~al.}, \bibinfo{journal}{Science}
      \textbf{\bibinfo{volume}{346}}, \bibinfo{pages}{1208} (\bibinfo{year}{2014}),
      ISSN \bibinfo{issn}{0036-8075},
      \urlprefix\url{http://www.sciencemag.org/lookup/doi/10.1126/science.1250366}.
    
    \bibitem[{\citenamefont{Luo et~al.}(2015)\citenamefont{Luo, Chen, Dai, Xu, and
      Thompson}}]{Luo2015}
    \bibinfo{author}{\bibfnamefont{Y.}~\bibnamefont{Luo}},
      \bibinfo{author}{\bibfnamefont{H.}~\bibnamefont{Chen}},
      \bibinfo{author}{\bibfnamefont{J.}~\bibnamefont{Dai}},
      \bibinfo{author}{\bibfnamefont{Z.-a.} \bibnamefont{Xu}}, \bibnamefont{and}
      \bibinfo{author}{\bibfnamefont{J.~D.} \bibnamefont{Thompson}},
      \bibinfo{journal}{Physical Review B} \textbf{\bibinfo{volume}{91}},
      \bibinfo{pages}{075130} (\bibinfo{year}{2015}), ISSN
      \bibinfo{issn}{1098-0121}, \eprint{1412.5449},
      \urlprefix\url{https://link.aps.org/doi/10.1103/PhysRevB.91.075130}.
    
    \bibitem[{\citenamefont{Hamidian et~al.}(2018)\citenamefont{Hamidian, Pirie,
      He, Soumyanarayanan, Yee, Kim, Rosa, Thompson, Fisk, Morr
      et~al.}}]{Hamidian2018}
    \bibinfo{author}{\bibfnamefont{M.}~\bibnamefont{Hamidian}},
      \bibinfo{author}{\bibfnamefont{H.}~\bibnamefont{Pirie}},
      \bibinfo{author}{\bibfnamefont{Y.}~\bibnamefont{He}},
      \bibinfo{author}{\bibfnamefont{A.}~\bibnamefont{Soumyanarayanan}},
      \bibinfo{author}{\bibfnamefont{M.}~\bibnamefont{Yee}},
      \bibinfo{author}{\bibfnamefont{D.-J.} \bibnamefont{Kim}},
      \bibinfo{author}{\bibfnamefont{P.}~\bibnamefont{Rosa}},
      \bibinfo{author}{\bibfnamefont{J.}~\bibnamefont{Thompson}},
      \bibinfo{author}{\bibfnamefont{Z.}~\bibnamefont{Fisk}},
      \bibinfo{author}{\bibfnamefont{D.}~\bibnamefont{Morr}}, \bibnamefont{et~al.},
      in \emph{\bibinfo{booktitle}{APS March Meeting}} (\bibinfo{year}{2018}).
    
    \bibitem[{\citenamefont{Alexandrov et~al.}(2015)\citenamefont{Alexandrov,
      Coleman, and Erten}}]{Alexandrov2015}
    \bibinfo{author}{\bibfnamefont{V.}~\bibnamefont{Alexandrov}},
      \bibinfo{author}{\bibfnamefont{P.}~\bibnamefont{Coleman}}, \bibnamefont{and}
      \bibinfo{author}{\bibfnamefont{O.}~\bibnamefont{Erten}},
      \bibinfo{journal}{Physical Review Letters} \textbf{\bibinfo{volume}{114}},
      \bibinfo{pages}{177202} (\bibinfo{year}{2015}), ISSN
      \bibinfo{issn}{0031-9007}, \eprint{1501.3031},
      \urlprefix\url{https://link.aps.org/doi/10.1103/PhysRevLett.114.177202}.
    
    \bibitem[{\citenamefont{Knolle and Cooper}(2015)}]{Knolle2015}
    \bibinfo{author}{\bibfnamefont{J.}~\bibnamefont{Knolle}} \bibnamefont{and}
      \bibinfo{author}{\bibfnamefont{N.~R.} \bibnamefont{Cooper}},
      \bibinfo{journal}{Physical Review Letters} \textbf{\bibinfo{volume}{115}},
      \bibinfo{pages}{146401} (\bibinfo{year}{2015}), ISSN
      \bibinfo{issn}{0031-9007}, \eprint{1507.00885},
      \urlprefix\url{https://link.aps.org/doi/10.1103/PhysRevLett.115.146401}.
    
    \bibitem[{\citenamefont{Erten et~al.}(2016)\citenamefont{Erten, Ghaemi, and
      Coleman}}]{Erten2016}
    \bibinfo{author}{\bibfnamefont{O.}~\bibnamefont{Erten}},
      \bibinfo{author}{\bibfnamefont{P.}~\bibnamefont{Ghaemi}}, \bibnamefont{and}
      \bibinfo{author}{\bibfnamefont{P.}~\bibnamefont{Coleman}},
      \bibinfo{journal}{Physical Review Letters} \textbf{\bibinfo{volume}{116}},
      \bibinfo{pages}{046403} (\bibinfo{year}{2016}), ISSN
      \bibinfo{issn}{0031-9007}, \eprint{1510.02313},
      \urlprefix\url{https://link.aps.org/doi/10.1103/PhysRevLett.116.046403}.
    
    \bibitem[{\citenamefont{Pal}(2018)}]{Pal2018}
    \bibinfo{author}{\bibfnamefont{H.~K.} \bibnamefont{Pal}}
      (\bibinfo{year}{2018}), \eprint{1801.05976v1}.
    
    \bibitem[{\citenamefont{Shen and Fu}(2018)}]{Shen2018}
    \bibinfo{author}{\bibfnamefont{H.}~\bibnamefont{Shen}} \bibnamefont{and}
      \bibinfo{author}{\bibfnamefont{L.}~\bibnamefont{Fu}},
      \textbf{\bibinfo{volume}{0}} (\bibinfo{year}{2018}), \eprint{1802.03023},
      \urlprefix\url{http://arxiv.org/abs/1802.03023}.
    
    \bibitem[{\citenamefont{Larson and Gordon}(1967)}]{Larson1967}
    \bibinfo{author}{\bibfnamefont{C.~O.} \bibnamefont{Larson}} \bibnamefont{and}
      \bibinfo{author}{\bibfnamefont{W.~L.} \bibnamefont{Gordon}},
      \bibinfo{journal}{Physical Review} \textbf{\bibinfo{volume}{156}},
      \bibinfo{pages}{703} (\bibinfo{year}{1967}), ISSN \bibinfo{issn}{0031-899X},
      \urlprefix\url{https://link.aps.org/doi/10.1103/PhysRev.156.703}.
    
    \bibitem[{\citenamefont{Onsager}(1952)}]{Onsager1952}
    \bibinfo{author}{\bibfnamefont{L.}~\bibnamefont{Onsager}},
      \bibinfo{journal}{Philosophical Magazine} \textbf{\bibinfo{volume}{43}},
      \bibinfo{pages}{1006} (\bibinfo{year}{1952}), ISSN \bibinfo{issn}{1941-5982},
      \urlprefix\url{http://www.tandfonline.com/doi/abs/10.1080/14786440908521019}.
    
    \bibitem[{\citenamefont{Wolgast et~al.}(2015)\citenamefont{Wolgast, Eo,
      {\"{O}}zt{\"{u}}rk, Li, Xiang, Tinsman, Asaba, Lawson, Yu, Allen
      et~al.}}]{Wolgast2015}
    \bibinfo{author}{\bibfnamefont{S.}~\bibnamefont{Wolgast}},
      \bibinfo{author}{\bibfnamefont{Y.~S.} \bibnamefont{Eo}},
      \bibinfo{author}{\bibfnamefont{T.}~\bibnamefont{{\"{O}}zt{\"{u}}rk}},
      \bibinfo{author}{\bibfnamefont{G.}~\bibnamefont{Li}},
      \bibinfo{author}{\bibfnamefont{Z.}~\bibnamefont{Xiang}},
      \bibinfo{author}{\bibfnamefont{C.}~\bibnamefont{Tinsman}},
      \bibinfo{author}{\bibfnamefont{T.}~\bibnamefont{Asaba}},
      \bibinfo{author}{\bibfnamefont{B.}~\bibnamefont{Lawson}},
      \bibinfo{author}{\bibfnamefont{F.}~\bibnamefont{Yu}},
      \bibinfo{author}{\bibfnamefont{J.~W.} \bibnamefont{Allen}},
      \bibnamefont{et~al.}, \bibinfo{journal}{Physical Review B}
      \textbf{\bibinfo{volume}{92}}, \bibinfo{pages}{115110}
      (\bibinfo{year}{2015}), ISSN \bibinfo{issn}{1098-0121}, \eprint{1409.8199},
      \urlprefix\url{https://link.aps.org/doi/10.1103/PhysRevB.92.115110}.
    
    \bibitem[{\citenamefont{Gunnersen}(1957)}]{Gunnersen1957}
    \bibinfo{author}{\bibfnamefont{E.~M.} \bibnamefont{Gunnersen}},
      \bibinfo{journal}{Philosophical Transactions of the Royal Society A:
      Mathematical, Physical and Engineering Sciences}
      \textbf{\bibinfo{volume}{249}}, \bibinfo{pages}{299} (\bibinfo{year}{1957}),
      ISSN \bibinfo{issn}{1364-503X},
      \urlprefix\url{http://rsta.royalsocietypublishing.org/cgi/doi/10.1098/rsta.1957.0001}.
    
    \bibitem[{\citenamefont{Caplan and Chanin}(1965)}]{Caplan1965}
    \bibinfo{author}{\bibfnamefont{S.}~\bibnamefont{Caplan}} \bibnamefont{and}
      \bibinfo{author}{\bibfnamefont{G.}~\bibnamefont{Chanin}},
      \bibinfo{journal}{Physical Review} \textbf{\bibinfo{volume}{138}},
      \bibinfo{pages}{A1428} (\bibinfo{year}{1965}), ISSN
      \bibinfo{issn}{0031-899X},
      \urlprefix\url{https://link.aps.org/doi/10.1103/PhysRev.138.A1428}.
    
    \bibitem[{\citenamefont{Phelan et~al.}(2016)\citenamefont{Phelan, Koohpayeh,
      Cottingham, Tutmaher, Leiner, Lumsden, Lavelle, Wang, Hoffmann, Siegler
      et~al.}}]{Phelan2016a}
    \bibinfo{author}{\bibfnamefont{W.~A.} \bibnamefont{Phelan}},
      \bibinfo{author}{\bibfnamefont{S.~M.} \bibnamefont{Koohpayeh}},
      \bibinfo{author}{\bibfnamefont{P.}~\bibnamefont{Cottingham}},
      \bibinfo{author}{\bibfnamefont{J.~A.} \bibnamefont{Tutmaher}},
      \bibinfo{author}{\bibfnamefont{J.~C.} \bibnamefont{Leiner}},
      \bibinfo{author}{\bibfnamefont{M.~D.} \bibnamefont{Lumsden}},
      \bibinfo{author}{\bibfnamefont{C.~M.} \bibnamefont{Lavelle}},
      \bibinfo{author}{\bibfnamefont{X.~P.} \bibnamefont{Wang}},
      \bibinfo{author}{\bibfnamefont{C.}~\bibnamefont{Hoffmann}},
      \bibinfo{author}{\bibfnamefont{M.~A.} \bibnamefont{Siegler}},
      \bibnamefont{et~al.}, \bibinfo{journal}{Scientific Reports}
      \textbf{\bibinfo{volume}{6}}, \bibinfo{pages}{20860} (\bibinfo{year}{2016}),
      ISSN \bibinfo{issn}{2045-2322},
      \urlprefix\url{http://www.nature.com/articles/srep20860}.
    
    \bibitem[{\citenamefont{Valentine et~al.}(2016)\citenamefont{Valentine,
      Koohpayeh, Phelan, McQueen, Rosa, Fisk, and Drichko}}]{Valentine2016}
    \bibinfo{author}{\bibfnamefont{M.~E.} \bibnamefont{Valentine}},
      \bibinfo{author}{\bibfnamefont{S.}~\bibnamefont{Koohpayeh}},
      \bibinfo{author}{\bibfnamefont{W.~A.} \bibnamefont{Phelan}},
      \bibinfo{author}{\bibfnamefont{T.~M.} \bibnamefont{McQueen}},
      \bibinfo{author}{\bibfnamefont{P.~F.~S.} \bibnamefont{Rosa}},
      \bibinfo{author}{\bibfnamefont{Z.}~\bibnamefont{Fisk}}, \bibnamefont{and}
      \bibinfo{author}{\bibfnamefont{N.}~\bibnamefont{Drichko}},
      \bibinfo{journal}{Physical Review B} \textbf{\bibinfo{volume}{94}},
      \bibinfo{pages}{075102} (\bibinfo{year}{2016}), ISSN
      \bibinfo{issn}{2469-9950}, \eprint{1601.02694},
      \urlprefix\url{https://link.aps.org/doi/10.1103/PhysRevB.94.075102}.
    
    \bibitem[{\citenamefont{Eo et~al.}(2018{\natexlab{b}})\citenamefont{Eo, Lucien,
      Rakoski, Mihaliov, Kurdak, Hatnean, and Balakrishnan}}]{YSEoMM2018}
    \bibinfo{author}{\bibfnamefont{Y.~S.} \bibnamefont{Eo}},
      \bibinfo{author}{\bibfnamefont{J.}~\bibnamefont{Lucien}},
      \bibinfo{author}{\bibfnamefont{A.}~\bibnamefont{Rakoski}},
      \bibinfo{author}{\bibfnamefont{D.}~\bibnamefont{Mihaliov}},
      \bibinfo{author}{\bibfnamefont{C.}~\bibnamefont{Kurdak}},
      \bibinfo{author}{\bibfnamefont{M.~C.} \bibnamefont{Hatnean}},
      \bibnamefont{and}
      \bibinfo{author}{\bibfnamefont{G.}~\bibnamefont{Balakrishnan}}, in
      \emph{\bibinfo{booktitle}{APS March Meeting}}
      (\bibinfo{year}{2018}{\natexlab{b}}).
    
    \end{thebibliography}
\end{document}